\documentclass[12pt]{article}
\usepackage{graphicx,a4} 
\usepackage{amssymb}
\usepackage{latexsym}
\usepackage{epsfig}

\def\lsim{\mathrel{\lower4pt\hbox{$\sim$}}
\hskip-12.5pt\raise1.6pt\hbox{$<$}\;}

\def\gsim{\mathrel{\lower4pt\hbox{$\sim$}}
\hskip-12.5pt\raise1.6pt\hbox{$>$}\;}

\begin{document}

\newcommand{\preprintno}[1]
{\vspace{-2cm}{\normalsize\begin{flushright}#1\end{flushright}}\vspace{1cm}}

\title{\preprintno{{\bf ULB-TH/03-06}}
Radion Assisted Gauge Inflation}
\author{{M. Fairbairn\thanks{E-mail: mfairbai@ulb.ac.be}}, L. Lopez Honorez\thanks{E-mail: llopezho@ulb.ac.be} and
{M.H.G. Tytgat\thanks{E-mail: mtytgat@ulb.ac.be}}\\
{\em Service de Physique Th\'eorique, CP225}\\
{\em Universit\'e Libre de Bruxelles}\\
{\em B-1050 Brussels, Belgium}}

\date{18th February 2003}

\maketitle
\begin{abstract}
We propose an extension to the recently proposed extranatural or gauge inflation scenario in which the radius modulus field around which the Wilson loop is wrapped assists inflation as it shrinks.  We discuss how this might lead to more generic initial conditions for inflation. 
\end{abstract}

Inflation is a compelling paradigm which solves many problems in cosmology whilst providing the seeds for the formation of structure \cite{Guth:1980zm}.  However in field theory there are no convincing candidates with the peculiar properties that the inflaton field must obey. Briefly the inflaton potential must be flat and yet provide enough potential energy to drive inflation.  One invariably requires some fine-tuning for both of these conditions to be satisfied at the same time. For instance, in the simplest version of chaotic inflation with a quadratic potential, inflation requires a trans-Planckian initial inflaton configuration \cite{Linde:gd}. In this regime, the effect of higher dimensional operators cannot be assumed to be negligible, which calls into question the naturalness of the theory.  This issue can be ameliorated to some extent if the inflaton potential is protected by symmetries, the most popular being supersymmetry. 
However, as pointed out in \cite{ed} and recently re-emphasised by Arkani-Hamed {\em et al} \cite{Arkani-Hamed:2003mz} and Kaplan and Weiner \cite{Kaplan:2003aj}, in the presence of gravity, supersymmetry alone can not explain the flatness of the inflaton potential. 

Another popular idea is that of natural inflation, in which the inflaton is the pseudo-Goldstone mode of a global symmetry, spontaneously broken at some scale $f$ and subsequently explicitly broken at a scale $\Lambda \ll f$ \cite{Freese:1990rb}. At low energies $\lsim \Lambda$, the inflaton effective Lagrangian has an axion-like form
\begin{equation}
\label{axion}
{\cal L} = {1\over 2} \partial_\mu \phi \partial^\mu \phi + \Lambda^4 \cos(\phi/f)
\end{equation}
The problem is that the conditions for a damped equation of motion and hence a slowly rolling field
\begin{equation}
\epsilon = {M_{pl}^2\over 2} \left({V^\prime \over V}\right)^2 \lsim 1\qquad\qquad  \left|\eta\right| = \left| M_{pl}^2 {V^{\prime\prime}\over V}\right| \lsim 1
\label{slowroll}
\end{equation}
impose the condition
$$
f^2 \gg M_{pl}^2
$$
which is again not natural from the field theory point of view. 

An interesting variation on this idea, dubbed extranatural or gauge inflation,  which appealingly overcomes this shortcoming  has been recently proposed by Arkani-Hamed {\em et al} \cite{Arkani-Hamed:2003wu,Arkani-Hamed:2003mz} and Kaplan and Weiner \cite{Kaplan:2003aj}. These authors have identified the inflaton with the phase $\theta$ of a Wilson loop wrapped around a compact dimension in a higher dimensional gauge theory. In the simplest version of this idea, with one extra dimension of radius $R$
$$
e^{i \theta} = e^{i \oint dx^5 A_5}.
$$
In the gauge in which the fifth component of the gauge field is constant along the extra dimension, $\theta = 2 \pi R A_5$.
 The potential for $\theta$ is flat at tree-level but at one-loop it takes a form similar to the axion potential (\ref{axion}),
\begin{equation}
\label{wilson}
{\cal L} = {1\over 2} {1\over g_5^2 2 \pi R} \partial_\mu \theta \partial^\mu\theta  + {c\over R^4} \cos(q \theta) + \ldots
\end{equation}
where $c = {\cal O}(1)$, $q$ is related to the charge of the particles that couple to the gauge field\footnote{For a  pure $SU(N_c)$ gauge theory $N\equiv N_c$ and the  the topological centre symmetry $Z(N_c)$ is spontaneously broken, similar to the weakly coupled phase of gauge theories at finite temperature \cite{Gross:1980br}. In the presence of other charged fields, the gauge symmetry is generically spontaneously broken.\cite{Hosotani:1983xw,Hatanaka:1998yp}} and the dots stand for sub-leading corrections.
The combination of five-dimensional parameters
$$
f^2 = {1\over g_5^2 2 \pi R}
$$
defines the scale of symmetry breaking, while $\Lambda \sim 1/R$. The slow-roll conditions still require $f^2 \gg M_{pl}^2$ but now this is merely the requirement for weak coupling in the five-dimensional gauge theory,
$$
g_5^2 2 \pi R M_{pl}^2 \ll 1
$$
As discussed in \cite{Arkani-Hamed:2003wu}, satisfying the slow-roll conditions whilst obtaining density perturbations of a magnitude consistent with the COBE observations typically requires
$$
R \approx 10 M_{pl}^{-1} \sim M_5^{-1}
$$
and 
$$
g_5^2 M_5\approx 10^{-3}
$$
where $M_5$ is the five-dimensional Planck constant, $M_{pl}^2 = 2 \pi R M_5^3$. 

The main benefit of this scenario is that the inflaton potential is protected by a gauge symmetry. Locality in five-dimensions cuts-off the potential at the scale $1/R$ and protects it from uncontrollable higher dimensional operators. Furthermore, non-perturbative quantum gravity corrections are expected to be exponentially suppressed provided $R M_5 \gg 1$, which is only marginally satisfied with only one extra dimension. It is however easy to construct  hybrid inflation versions of this scenario by considering more dimensions with different radii (see \cite{Arkani-Hamed:2003mz,Kaplan:2003aj} for details). In this work however, we would like to advocate a slightly different extension that we believe is a very natural variation of the idea of gauge inflation. 

To motivate our discussion we recall another attractive feature of the old idea of natural inflation. Since the inflaton in this model is a cyclic variable, the problem of initial conditions for inflation is neatly solved if the inflaton field can take random values at some earlier time \cite{Freese:1990rb}. Drawing on the analogy with the axion effective potential, above the energy scale $\Lambda$ the potential appears effectively flat and, say from thermal fluctuations, the  rms of the cyclic field is
$$
{\left\langle\vert{ \phi/  f}\vert\right\rangle} = \sqrt{\pi/ 3} \sim 1
$$
As the universe cools down, the probability that the inflaton is initially displaced from its minimum homogeneously across one horizon by an amount sufficient to begin inflation can be quite large \cite{Freese:1990rb}.
Going back to the gauge version of natural inflation, we can envisage a similar situation in which the phase of the Wilson loop initially takes random values and only later on relaxes to the minimum of its effective potential. 
One might expect the energy scale at which the effective potential is negligible to be of order $R^{-1}$, but this is too naive. Laine and Korthals-Altes have studied the finite temperature phase diagram of non-abelian gauge theories in $D=3+1$ and $D=4+1$ with one compact dimension and have argued that the phase transition actually takes place through the coalition of domain walls at a temperature (for $D=4+1$) of order \cite{KorthalsAltes:2001et}
$$
T_c \sim  g_5^{-2}
$$
and this estimate is supported by lattice simulations \cite{Farakos:2002zb}.
For the problem at hand however, this temperature is far above the Planck scale and so does not make much sense.  A more natural way of getting a flatter potential  for $\theta$ arises if the size of the compact dimension can change. Inspection of the effective Lagrangian (\ref{wilson}) reveals that if the radius $R$ is {\em larger} at some earlier time there are two effects:  a) the potential at this time is flatter and b) large gradients in the inflaton field cost less energy.\footnote{Note that, by definition, $\theta$ is independent of the metric and thus is invariant under a rescaling of the fifth dimension.} These features go in the right direction to help provide the initial conditions for inflation. We are thus naturally led to relax the condition of a fixed extra dimension and to take into account the effect of a varying modulus field upon the dynamics. 

Before proceeding, we have to face several issues. First we would like to work at the level of an effective $D=3+1$ field theory as much as possible. This implies that we should only consider scales large compared to the radius of the extra dimension. Secondly, if the extra dimension is dynamical at early times, we need a mechanism to stabilise it at late times.  This is a notoriously difficult problem for higher dimensional theories for which there is no definitive solution.
 
Finally, in principle we should compute the quantum effective action for the coupled radion and Wilson loop system.  However, in this paper we will try to be pragmatic and write down a simple model which hopefully mimics the impact of a varying radius on the dynamics of the inflaton field. Our toy model is given by an effective potential of the form
 
\begin{equation}
{\cal L} = {1\over 2} \partial_\mu \chi \partial^\mu \chi + {1\over 2} f^2 \partial_\mu \theta\partial^\mu \theta
 - \; {1\over 2} \chi^4 {\phi^2\over f^2} - {\lambda\over 4} \left(\chi^2 - \chi_0^2\right)^2.
\end{equation}
where $\phi = \theta f$.

Let us comment on the different terms in this effective action. Firstly, we have introduced a radion field $\chi$,  related to the fifth diagonal component of the metric $G_{55}$, $\chi \sim 1/{2\pi R}$, and performed a Weyl rescaling so as to work in the Einstein frame. The $D=3+1$ Planck mass is thus fixed, $M_{pl}^2 = 2 \pi R_0 M_5^3$. The kinetic term for the radion is taken to be canonical for simplicity. Second, we have set $f^2 = (g_5^2 2 \pi R_0)^{-1}$ to a  constant in the kinetic term for $\theta$.  For this assumption to be meaningful, we will have to ensure later that the expectation of the radion field $\chi$ does not change much over the period of inflation.  Then we have assumed that the radion-inflaton coupling is analogous to the one valid for a frozen radius (\ref{wilson}) and expanded the cosine keeping only the leading quadratic term.  Finally, for the sake of illustration, the radion potential is assumed to take a simple quartic form, with a minimum at $R = R_0$.

\begin{figure}
\begin{center}
\epsfig{file=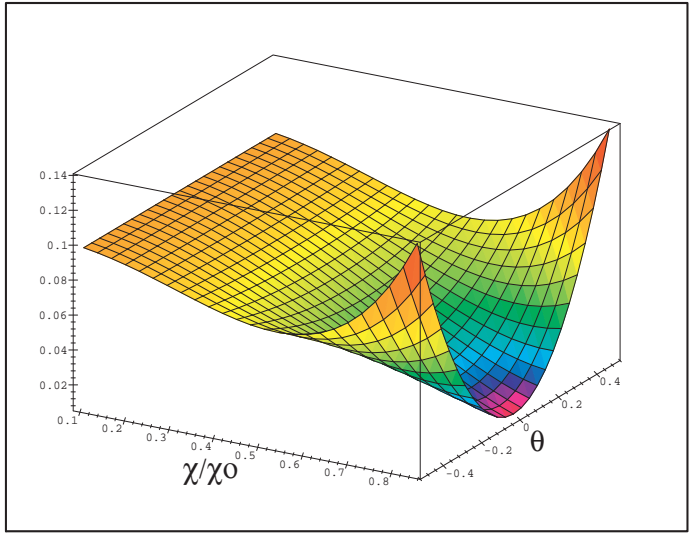,width=8cm,height=10cm,angle=270}
\caption{\label{plot1}Schematic plot of the radion-inflaton effective potential.
}
\end{center}
\end{figure}

\begin{figure}
\begin{center}
\epsfig{file=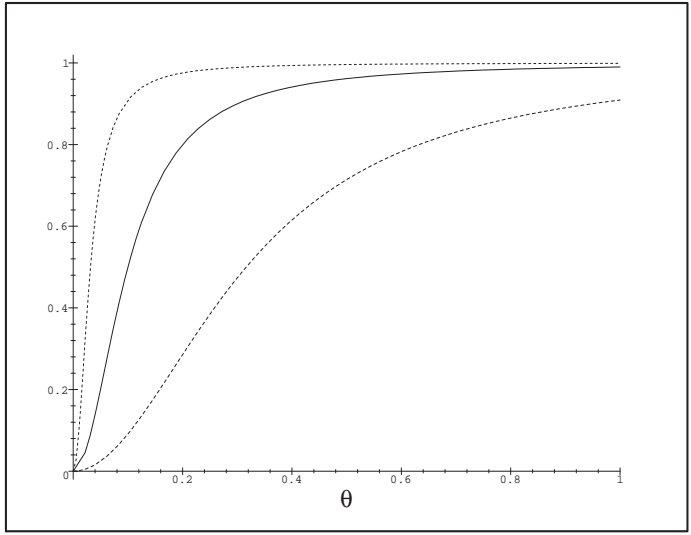,width=8cm,height=10cm,angle=270}
\caption{\label{flatness}Plot of the radion-inflaton effective potential for $\lambda=10^{-1},10^{-2}$ and $10^{-3}$.  The potentials, which have been normalised for comparison, become flatter for smaller $\lambda$.}
\end{center}
\end{figure}

Having simplified the problem to its extreme (hopefully without killing the underlying physics), we are left with two potentially interesting regimes.  First, if the radion self-coupling is ``strong'' (to be defined soon) the radius is essentially frozen, $\langle \chi \rangle = \chi_0$ and we get back to the effective action (\ref{wilson}) already discussed by Arkani-Hamed {\em et al} \cite{Arkani-Hamed:2003wu}. The alternative is also, we believe, quite interesting. For $\phi=0$, if the universe starts with an initially large  extra dimension, the  $\chi$ field rolls from a small value of $\chi$ toward $\chi_0$, If the  inflaton field $\phi$ is non-zero however, it induces an effective quartic coupling for the radion which can  compete with the radion self-coupling. For $\phi/f \gsim \lambda$, with $\lambda \ll 1$ there is a throat along which $\chi \lsim \chi_0$ and also $V \sim V_0$. (See figure \ref{plot1}.) As we shall see, the inflaton field can be made to slowly roll along this throat.  This scenario might at first look  analogous to hybrid inflation or one of its avatars but it is slightly different because the $\chi$ field never becomes tachyonic. Rather, the inflaton is a tachyon, but with an initially small negative mass square provided $\phi/f \gsim \lambda$.\footnote{Although we are not aware of any reference, it is plausible that such an effective potential for inflation has been discussed before, but presumably not with the same underlying physics in mind. }
To see these features, we differentiate the potential 
\begin{equation}
\label{effpot}
V(\chi, \phi) = {1\over 4} \chi^4 \phi^2/f^2 + {\lambda\over 4}(\chi^2 - \chi_0^2)^2
\end{equation}
with respect to $\chi$ in order to obtain the minimum of the effective potential for different values of $\phi$ and hence the expectation value of $\chi$
$$
\langle\chi^2 \rangle = {\chi_0^2 \over 1 + \phi^2/\lambda f^2}.
$$ 
Substituting this back into (\ref{effpot}) we obtain a simplified parametric form of the potential energy in terms of $\phi$ alone,
\begin{equation}
\label{inflaton}
\tilde V(\phi) = {\lambda \over 4}{ \chi_0^4 \phi^2\over f^2 \lambda + \phi^2}
\end{equation}
which we will take as our effective inflaton field.  When $\phi^2/f^2 \gsim 4 \lambda$ the potential is nearly flat and has negative curvature.  The potential becomes steeper for $\lambda \lsim \phi^2/f^2 \lsim 4 \lambda$ and finally has a positive curvature around the origin. (See figure \ref{flatness}.)  As $\phi$ decreases from $f$ towards the end of the flat region of the potential, the fractional variation in $\chi$ is  ${\cal O}( \sqrt\lambda f/\phi)$, after which $\langle\chi\rangle$ moves rapidly toward $\chi_0$. Since we will see that we require $\lambda\ll 1$, this vindicates our not taking into account the effect of the variation of $\chi$ upon the kinetic term for $\theta$ during inflation.

Now we are able to do the usual inflationary calculations.  The slow-roll conditions (\ref{slowroll}) take the form
$$
\epsilon = 2 {M_{pl}^2\over \phi^2} {1\over (1 + \phi^2/\lambda f^2)^2} \lsim 1
$$
and
$$
\eta = 2{M_{pl}^2\over \phi^2} {(1 - 3 \phi^2/\lambda f^2)\over (1+ \phi^2/\lambda f^2)^2} \lsim 1
$$
such that $|\eta|\gg\epsilon$ and we do not expect gravitational waves to be produced.

It is interesting to compare these expressions to those of the frozen radion scenario. The latter corresponds to the limit of large radion self-coupling $\lambda \gg \phi^2/f^2$ and slow-roll inflation requires typically that $\phi^2 \gg M_{pl}^2$. In the present case however the effective potential has a plateau for $\phi^2/f^2 \gsim \lambda$. In this regime, the slow-roll conditions \cite{Liddle:cg} become
$$
\epsilon \approx 2 \lambda^2 {M_{pl}^2 f^4 \over \phi^6}
\label{epsilon}
$$
and
\begin{equation}
\eta \approx - 6 \lambda {M_{pl}^2 f^2\over \phi^4}.
\label{eta}
\end{equation}
Taking for instance the conservative value $f \sim M_{pl}$, inflation stops for $\phi\sim\lambda^{1/4} M_{pl}$ which is much lower than $M_{pl}$ for small values of the coupling $\lambda$. The spectral index $n_s$ is obtained from the slow roll parameters in the usual way
$$
n_s = 1 - 6 \epsilon + 2\eta \approx 1 + 2 \eta = 1 - 12 \lambda {M_{pl}^2 f^2\over \phi^4} 
$$
and thus is less than one, consistent with the newly released WMAP observations \cite{Peiris:2003ff}.  It is also quite easy to compute the number of e-folds of inflation,
\begin{eqnarray}
N &\approx& {1\over M_{pl}^2} \int_{\phi_{end}}^{\phi_i} {\tilde V\over \tilde V^\prime} d\phi \nonumber\\
&= & {1\over 4 M_{pl}^2} \left [{\phi^2} \left(1 + \phi^2/4 \lambda f^2\right)\right ]_{\phi_{end}}^{\phi_i}\\
&\approx& {f^2\over 16 M_{pl}^2 \lambda}
\label{efolds}
\end{eqnarray}
where we have taken the initial value of the $\phi$ field as $\phi_i \lsim f$.  Then, for $f \sim M_{pl}$ we require $\lambda \lsim 10^{-(2-3)}$ in order to get more than 60 e-folds.
Finally, the normalisation of the magnitude of the perturbations to COBE requires
$$
{\tilde V^{3/2}\over M_{pl}^3 \tilde V^\prime} \approx 5.2 \times 10^{-4}$$
where
$$
{\tilde V^{3/2}\over M_{pl}^3 \tilde V^\prime} \approx {\chi_0^2 \phi^2 \over 4 M_{pl}^3  f} (1 + \phi^2/\lambda f^2)^{1/2} \approx {\chi_0^2 \phi^3\over 4 \sqrt{\lambda} M_{pl}^3 f^2}
$$ 
Typically, one obtains slow roll, a flat spectrum, enough e-folds and suitable perturbations with parameters of the order of
$$
\lambda \sim 10^{-3}, \qquad f \sim \phi_i \sim M_{pl} \qquad 2 \pi R_0 \sim 10^{2} M_{pl}^{-1}$$
In turn, these conditions imply that the five-dimensional parameters satisfy
\begin{equation}
\label{g5}
g_5^2 \approx 10^{-2} M_{pl}^{-1}
\end{equation}
corresponding to
\begin{equation}
\label{R0}
2 \pi R_0 \approx 10^{4/3} M_{5}^{-1}
\end{equation}
which also shows that the energy density during inflation is far below the 5D Planckian energy density.\footnote{By the same token during slow-roll $R \sim 10^{-2} H^{-1}$ where the inverse Hubble parameter $H^{-1}$ gives the size of the Horizon. That $R \ll H^{-1}$ always is a necessary condition for the  consistency of the $D=3+1$ effective theory formulation.} Finally, the radion mass turns out to be of order
$$
M_{\chi} \sim 10^{-4} M_{pl} \sim M_{\phi}
$$

It is tempting to  see if this picture is compatible with string theory. However, at this stage, this is of limited interest since we do not know how to relate the radion potential to string theory. Nevertheless,  following \cite{Arkani-Hamed:2003wu}, let us assume that $M_{pl}^2=g_s^{-2}(2\pi R)^6 l_s^{-8}$ and $f=g_s^{-1}(2\pi R)^2 l_s^{-3}$ where $g_s$ and $l_s$ are the string coupling and length respectively.  This scaling corresponds to an isotropic compactification of the higher dimensions with the Wilson loop running around one of them.  
Since 
$
{f/M_{pl}} = {l_s/ 2 \pi R_0}
$
the inequality $f \lsim M_{pl}$ ensures that the size of the compact dimensions is bigger than the string length and we not have to go to the T-dual picture.   
Then, from (\ref{R0}), setting $2 \pi R \approx l_s$ (or $f \approx M_{pl}$) gives
$
g_s \sim 10^{-2}
$
. That $\lambda \sim g_s^2$ is presumably an accident.

\bigskip 

In conclusion, we have investigated the effect of making the extra dimension dynamical in extranatural or gauge inflation.  In these theories the inflationary potential is the effective potential generated when a non zero Wilson loop is switched on in the compactified directions in the presence of charged particles.  The angular variable $\theta$ of the Wilson loop in such models becomes the inflaton.

In order for there to be good initial conditions for inflation, one would like the universe to start in a regime where the fluctuations of the inflaton field $\theta$ are large enough to be quite evenly distributed from 0 to 2$\pi$.  However, inflation places tight constraints on the parameters of the potential and the typical energy scale at which the energy of the inflationary potential becomes negligible is trans-Planckian. 
In gauge inflation, the height of such a potential is controlled by the inverse of the compactification scale.  We have therefore argued that if the extra dimension starts large, the field $\theta$ might more naturally spread out during this early epoch and obtain suitable initial conditions for inflation.  We have gone on to show that the evolution of the extra dimension during inflation changes the shape of the effective inflationary potential and leads one to different constraints upon the model parameters.

At this early stage in our analysis we are aware that we have made many approximations and the scenario described above should be considered at best a toy model.  However, although improvements would be most welcome, we believe that our calculations should convey at least the flavour of the underlying physics accurately. 

\section*{Acknowledgements}
We are grateful for conversations with Jean Orloff.

\end{document}